\documentstyle[prl,aps,floats,epsf]{revtex}
\begin{document}
\draft

\twocolumn[\hsize\textwidth\columnwidth\hsize\csname
@twocolumnfalse\endcsname

\title{Proper ferroelastic phase transitions in thin epitaxial films with
symmetry-conserving and symmetry-breaking misfit strains}
\author{A.M. Bratkovsky$^{1}$ and A.P. Levanyuk$^{1,2}$}
\address{
$^{1}$Hewlett-Packard Laboratories, 
1501 Page Mill Road, Palo Alto, California 94304\\
$^{2}$Departamento de F\'{i}sica de la Materia Condensada,\\
C-III, Universidad Aut\'{o}noma de Madrid, 28049 Madrid, Spain}
\date{ March 31, 2001 }
\maketitle

\begin{abstract}

We study how the ferroelastic domain structure sets in in an epitaxial film of
a material with second order proper ferroelastic transition. The domain
structures considered are similar to either $a_{1}/a_{2}/a_{1}/a_{2}$ or 
$c/a/c/a$ structures in perovskite ferroelectrics. If the ``extrinsic"
misfit strain,  
not associated with the transition, does not break the symmetry of 
the high-temperature
phase, the phase transition in the film occurs at somewhat lower temperature
compared to the bulk. The loss of stability then occurs with respect to a sinusoidal
strain wave, which evolves into the domain structure with practically 
the same geometry and
approximately the same period. In the presence of the symmetry-breaking
component of the misfit strain (``extrinsic" misfit)
the character of the phase transition is
qualitatively different. In this case it is a {\em topological} transition between
single-domain and multi-domain states, which starts from a low density of 
the domain walls.


\end{abstract}
\vskip 2pc ] % end \twocolumn[...]

\narrowtext

\section{Introduction}

Symmetry breaking spontaneous strains appear frequently during phase
transitions in various materials, and. these materials behave as either
proper or improper ferroelastics. Ferroelastic phase transitions in thin
films usually lead to formation of domain structures in order to accommodate
so-called ``misfit stresses'', which appear at the transition or may already
exist in the high-symmetry phase because of the difference in the unit cell
parameters of a film and a substrate \cite{Roitburd}. Currently there is
much interest in the ferroelastic domain structures in thin films of
``pure'' ferroelastics (see, e.g. \cite{Streifer,Seifert}) and perovskite
ferroelectrics, which are improper ferroelastics (see references in Ref.\cite
{Roytburd98}).

The continuous-medium description applies to these domain patterns, since
usually the thickness of the films is by two to three orders of magnitude
larger than the unit cell parameter. One can also expect that the results of
this approach will be qualitatively valid even for very thin films. It
should be noted that a rigorous continuous medium treatment of ferroelastic
domain structure is\ a problem of both non-linear and non-local elasticity
in inhomogeneous medium (e.g. film on a substrate) and presents tremendous
mathematical difficulties. Even after being reduced to a linear local
elasticity problem in a homogeneous medium (i.e. for the same elastic
constants of the film and the substrate) it is usually approached with a
combination of approximate and numerical methods (see, e.g. \cite
{Roytburd91,Roytburd93,Pompe93,Pertsev95,Sridhar96,Roytburd98}). Recently we
have treated two simple yet generic examples of domain structures, which
correspond to either $a_{1}/a_{2}/a_{1}/a_{2}$ or $c/a/c/a$ domain patterns
in perovskite ferroelectrics\cite{BLxy1,BLca}. Unlike the previous authors,
we were able to obtain the analytical expressions for energies of the domain
structures. This allowed us to reproduce the available results, and obtain
the new ones. In the present paper we shall analyze the loss of stability of
the symmetric phase, which eventually leads to formation of the domain
pattern.

We shall start with the question of how the domain structures appear at the
phase transition\cite{Part}. More specifically, we suppose that there occurs
a second-order proper ferroelastic transition in a free sample,
characterized by a one-component order parameter. We consider first the case
where the substrate induces no misfit strain apart from that appearing at
the phase transition. If this misfit strain can be accommodated by a domain
structure, the domain pattern inevitably appears at such a transition. It
proceeds with the loss of stability of the symmetric phase with respect to a
``strain wave'', which then evolves into the domain structure. The process
is somewhat different for an improper ferroelastic: the domain structure
there may not appear immediately at the transition, since there a
single-domain state is metastable. Therefore, the domain structure may
develop there at lower temperatures. However, the structures are similar in
both proper and improper ferroelastics not very close to the critical point,
since they are mainly defined by the symmetry of the spontaneous strain.
Therefore, theoretical study of proper ferroelastic transitions in thin
epitaxial films is of interest on its own merits, and because it reveals the
general properties of domain structures, even when the strain is not a
``primary'' order parameter.

We find the form of the ``strain waves'', which may appear at the second
order transition. The problem is non-local but linear, that is why we cannot
find the {\em amplitude} of the ``frozen strain wave'' just below the
transition. However, the form and the period of the ``strain wave'' close to
the transition proves to be, in some cases, practically the same as far from
the transition, where this ``wave'' transforms into the domain structure.
Therefore, the important features of domain structures can be found within a
linear theory, and this makes the problem of the stability loss even more
interesting.

The misfit strain, not associated with the phase transition, if any, changes
the situation significantly if it contains the strain corresponding to the
order parameter. Then, the phase that has been symmetric in the free sample
is no longer symmetric in the epitaxial film. However, a continuous phase
transition may still take place. The homogeneous non-symmetric state, which
does not ``feel'' the lowering of temperature below the phase transition in
the free sample, may transform, at yet lower temperature, into an
inhomogeneous multidomain state. This transition is similar to
commensurate-incommensurate transition or the transition between homogeneous
and vortex states in type II superconductors in magnetic field. Such
transitions, while being continuous, are not accompanied by a stability
loss. The phase diagrams (including those in the plane of variables misfit
strain-film thickness) of transition between single- and multidomain states
were studied by several authors\cite{Roytburd93,Pompe93,Pertsev95}. Some
properties of the domain structure near this transition we have discussed
earlier \cite{BLca}. We briefly discuss this topic below in the context of
the present paper.

We shall consider the epitaxial film with the plane perpendicular to $z-$%
axis. The misfit stress, not associated with the phase transition (
``extrinsic'' misfit), contains two components of the strain tensor: $%
u_{xx}^{0}=\left| u_{yy}^{0}\right| ,$ with the unstrained state for the
free film being just above the phase transition. In Sec. 2 we analyze how
the domain structure sets in at a proper ferroelastic transition with $u_{xy}
$ the order parameter. We find the period of the ``strain wave'' appearing
at the transition and estimate the critical thickness of the film where the
ferroelectric phase exists. The extrinsic misfit is not important in this
case. In Sec. 3 we treat the case of $u_{xx}-u_{zz}$ as the order parameter
for the same film geometry, first for the case when the permanent misfit is
absent and then when it is present. The present results are summarized at
the end.

\section{Strain $u_{xy}$ as the order parameter}

Consider the substrate as an infinite isotropic medium with the shear
modulus $\mu $. The elastic moduli of the ferroelastic are supposed to be
the same as in the substrate, with the exception of the ``soft'' modulus
corresponding to the $u_{xy}$ component of strain, Fig.~1. One can guess the
important features of the ferroelastic domain structure by just considering
the symmetric phase. Indeed, the domain pattern evolves from a ``frozen
strain wave'' appearing at the transition. Given the macroscopic scale of
the wave, one expects that the corresponding strains are mainly the ``soft''
ones. This immediately provides the possible directions of the ``wave
front'', which corresponds to the possible orientations of the domain walls.
In the present case the front is perpendicular either to $x$- or $y$-axes,
since for all other directions the acoustic modes are not ``soft''. Indeed,
they involve not only the soft $u_{xy}$-component of the strain tensor but
also other (``hard'') components. Therefore, to study the stability loss we
may consider either the waves of the displacement vector component $u_{x}$,
propagating along the $y$-axis or swap the $y-$ and $x-$axes. Evidently,
each wave is characterized by the strain $2u_{xy}=\left( \partial
u_{x}/\partial y\right) +$ $\left( \partial u_{y}/\partial x\right) $ and
the rotation $2\Omega _{xy}=\left( \partial u_{x}/\partial y\right) -$ $%
\left( \partial u_{y}/\partial x\right) $ with $\left| \Omega _{xy}\right|
=\left| u_{xy}\right| .$ This elementary result illustrates the fact that a
ferroelastic domain structure involves both strains and rotations, as
discussed at some length in, e.g., Ref.\cite{Jacobs00}. The wave we consider
is confined within the film, therefore, we have to take into account the
dependence of $u_{x}$ (or $u_{y}$) on $z$.
\begin{figure}[t]
\epsfxsize=3.2in \epsffile{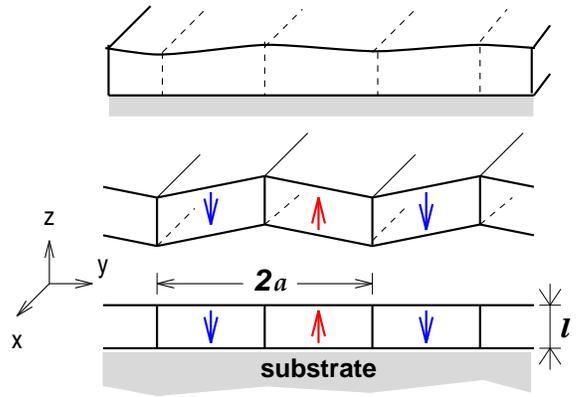}
\bigskip
\caption{Schematic of the $a_1/a_2/a_1/a_2$ ``strain wave'' (top
panel) and the domain structure (middle and bottom panels) in thin
epitaxial ferroelastic film with the spontaneous strain $u^0_{xy}$.
The symmetric phase looses its stability at some temperature $T_{c1}$
below the phase transition temperature in free film,  $T_c$, and forms the
``strain wave'' (top panel). 
Upon further cooling it transforms into the domain structure
with the period $2a$ (bottom panel) in the film of thickness $l$.
}
\label{fig:figxy}
\end{figure}

It is obvious from the above that we have to retain in the elastic (free)
energy the terms dependent on $u_{xy}$ and $u_{xz}$ (or $u_{xz}$) only.
Among the gradient terms there are, strictly speaking, those containing the
gradients of $\Omega _{ik}$\cite{Jacobs00}, but their account does not
affect the results and we shall omit them together with the gradients of the
``hard'' components of the strain tensor. Therefore, we choose the Landau
free energy in the form

\begin{equation}
F=\int dV\left[ 2Au_{xy}^{2}+2D\left( \nabla u_{xy}\right) ^{2}+\mu \left(
u_{ik}^{2}-2u_{xy}^{2}\right) \right]  \label{eq2.1}
\end{equation}
where $A=\alpha \left( T-T_{c}\right) $, with $\alpha ,D,\mu $ positive
constants. Recall that higher order terms are not needed for the analysis of
loss of stability of the symmetric phase. We obtain, therefore, only one
non-local equation of state:

\begin{equation}
\sigma _{xy}\equiv \frac{1}{2}\frac{\delta F}{\delta u_{xy}}=2(A-D\nabla
^{2})u_{xy}  \label{eq2.2}
\end{equation}

The stability loss corresponds to an appearance of a non-trivial solution of
the linearized equations of equilibrium. Indeed, just at the brink of
instability the system is in neutral equilibrium, if the non-linear terms
are not taken into account. We shall look for this non-trivial solution for
the $x$-component of the displacement vector

\begin{equation}
u_{x}\equiv u(y,z)  \label{eq2.3}
\end{equation}
It has been mentioned above that the same result is obtained when the $x$
and $y$ axes are interchanged.

Let us recall firstly that, while studying the equilibrium of an elastic
system, one has to consider separately the net strains, i.e. the homogeneous
part of the strains, and their inhomogeneous part. The point is that the
homogeneous part, which is described by six independent components of the
(mean) strain tensor defines the change of the volume and the shape of the
sample. The inhomogeneous part is better described by three independent
components of displacement vector, since the use of the spatially varying
strain components is usually less convenient as they should satisfy the
additional (elastic compatibility) conditions.

One can see immediately that in the present case there is no loss of
stability with respect to homogeneous deformation. Indeed, if our sample
(film plus substrate) has a homogeneous deformation $\left(
u_{xy}^{h}\right) ,$ the energy of this deformation would be 
\begin{equation}
F=\left( 2Al+2\mu L\right) \left( u_{xy}^{h}\right) ^{2},  \label{eq2.3a}
\end{equation}
where $L$ is the thickness of the substrate, and it would be infinite in the
present case, $L=\infty $.

The inhomogeneous part of the strain should satisfy the equation of local
equilibrium 
\begin{equation}
\frac{\partial \sigma _{xy}}{\partial y}+\frac{\partial \sigma _{xz}}{%
\partial z}=0.  \label{eq2.4}
\end{equation}
We shall apply the Fourier transform 
\begin{equation}
u(y,z)=\int u_{k}\left( z\right) \exp \left( iky\right) dk  \label{eq2.5}
\end{equation}
and look for the appearance of the non-trivial solution of the equation of
equilibrium (\ref{eq2.4})\ for a given wave vector $k$. We shall then
determine the $k$ where the stability sets in first (at highest
temperature). Evidently, this would be the point where the symmetric phase
looses its stability.

For the ferroelastic $\left( 0<z<l\right) $ one obtains 
\begin{equation}
\sigma _{xy}\left( k,z\right) =2A_{k}u_{xy}\left( k\right)
=ikA_{k}u_{k}\left( z\right) ,  \label{eq2.6}
\end{equation}
where $A_{k}=A+Dk^{2},$ and 
\begin{equation}
\sigma _{xz}\left( k,z\right) =\mu \frac{du_{k}}{dz}.  \label{eq2.7}
\end{equation}
Equation $\left( \text{\ref{eq2.4}}\right) $ in the film ($0<z<l)$ takes the
form 
\begin{equation}
\frac{d^{2}u_{k}}{dz^{2}}-\frac{A_{k}}{\mu }k^{2}u_{k}=0.  \label{eq2.8}
\end{equation}
For the substrate $\left( -\infty <z<0\right) $ one has 
\begin{equation}
\frac{d^{2}u_{k}}{dz^{2}}-k^{2}u_{k}=0.  \label{eq2.9}
\end{equation}
At the stress-free surface $\left( z=l\right) $ the condition $\sigma
_{xz}\left( k,l\right) =0$ should be satisfied, or, taking into account Eq. $%
\left( \ref{eq2.7}\right) ,$ 
\begin{equation}
\frac{du_{k}}{dz}\mid _{z=l}=0  \label{eq2.10}
\end{equation}
At the interface $\left( z=0\right) $ the displacement $u_{k}\left( z\right) 
$ and the stress $\sigma _{xz}\left( k,z\right) $ should be continuous, and
the stress should vanish at $z\rightarrow -\infty .$

Let us first consider the case of $A_{k}<0,$ which would correspond to a
loss stability of the symmetric phase. The solution of Eqs. (\ref{eq2.8}), (%
\ref{eq2.10}) is

\begin{equation}
u_{k}(z)=F\cos \chi k(z-l),  \label{eq2.15a}
\end{equation}
where $\chi ^{2}=-A\left( k\right) /\mu ,$ while for the substrate we have

\begin{equation}
u_{k}(z)=G\exp \left| k\right| z.  \label{2.15b}
\end{equation}
The conditions at the interface give the equation for the existence of
non-trivial solutions

\begin{equation}
\cot \chi kl=\chi .  \label{eq2.16}
\end{equation}
This equation has a solution for $\chi ^{2}>0$, while there is no solution
for $\chi ^{2}<0$, hence the loss of stability takes place for $A_{k}<0.$
For the region of interest we should obviously have $\chi \ll 1,$ as we
shall check later, and the approximate solution is

\begin{equation}
\chi \simeq \frac{\pi }{2kl},  \label{eq2.17}
\end{equation}
or 
\begin{equation}
\left| A\right| -Dk^{2}=\frac{\pi ^{2}\mu }{4k^{2}l^{2}}.  \label{eq2.18}
\end{equation}
The minimum value of $|A|=\left| A\right| _{c}$ corresponds to the highest
temperature where the instability sets in with $k=k_{m}:$ 
\begin{equation}
k_{m}=\frac{\pi ^{1/2}\mu ^{1/4}}{2^{1/2}D^{1/4}l^{1/2}}\sim \frac{1}{%
d_{at}^{1/2}l^{1/2}},  \label{eq2.19}
\end{equation}
where usually (not on very ''soft'' substrates) $\left( D/\mu \right)
^{1/2}\sim d_{at}$ with $d_{at}$ the interatomic distance (see, e.g.,\cite
{Strukov}). Therefore, $\chi _{m}\sim 1/k_{m}l\sim \sqrt{d_{at}/l}\ll 1$,
and this confirms our initial assumption that $\chi $ is small.

As a result, the loss of stability of the symmetric phase takes place at 
\begin{equation}
\left| A\right| _{c}=\pi \frac{D^{1/2}\mu ^{1/2}}{l},  \label{eq2.19a}
\end{equation}
or, equivalently, at critical temperature, which is below that of a free
sample: 
\begin{equation}
T=T_{c1}=T_{c}-\pi \frac{D^{1/2}\mu ^{1/2}}{\alpha l}.  \label{eq2.20}
\end{equation}
For {\em displacive} systems $\alpha \sim \mu /T_{at}$, where $T_{at}$ is
the characteristic ''atomic'' temperature (see, e.g.,\cite{Strukov}). One
then finds that the lowering of the phase transition temperature is
approximately 
\begin{equation}
T_{c}-T_{c1}\sim T_{at}\frac{d_{at}}{l}.  \label{eq2.21}
\end{equation}
Since usually $T_{at}\sim \left( 10^{2}-10^{3}\right) T_{c},$ one may expect
a complete suppression of the second order transition in films with
thickness of hundreds atomic layers, which is accompanied by the loss of
stability of the symmetric phase. In order-disorder systems the shift of $%
T_{c}$ is just ($10^{-1}-1)$K. Certainly, one has to use the actual values
of the coefficients $D,\alpha ,\mu $ for a specific substance to decide if
the second order transition is indeed suppressed in a given material.

The stability gets lost with respect to inhomogeneous strains $u_{xy},u_{xz}$%
. The form of the ``strain wave'' in the film is

\begin{eqnarray}
u_{xy} &=&C\sin \left( k_{m}y+\varphi \right) \sin \frac{\pi z}{2l},
\label{eq2.21a} \\
u_{xz} &=&C\frac{\pi }{2k_{m}l}\cos \left( k_{m}y+\varphi \right) \cos \frac{%
\pi z}{2l},
\end{eqnarray}
where $C$ is the constant, which remains undetermined within the linear
theory. We see that the strain component $u_{xy},$ which corresponds to the
spontaneous strain in a free sample, is maximal near the free surface of the
film, where it is easier to deform it. At the same time, the
``accompanying'' strain $u_{xz}$ is concentrated near the film-substrate
interface and is much smaller than the $u_{xy}$ strain for $l\gg d_{at},$
because $k_{m}l\gg 1.$ To find the amplitude $C$ of the ``strain wave'' one
has to take into account the non-linear terms in the equation of state. This
is a difficult technical problem, and it is not addressed here. We shall
just mention that at lower temperatures the sinusoidal structure transforms
into a ``domain-like'' structure with the amplitude close to the value of
the spontaneous strain in a free crystal. One can interchange $x$ and $y$ to
obtain an alternative variant of  the ``strain wave''.

One can check that an inhomogeneous sinusoidal structure appearing at a
second order phase transition because of the boundary conditions is a
precursor of a domain structure or, more precisely, it is the form of the
domain structure close to the phase transition, where the widths of the
domain walls and the domains themselves become comparable (see, e.g. \cite
{Chensky}). Indeed, the domain wall width, $W_{c},$ is 
\begin{equation}
W_{c,ins}=\sqrt{\frac{D}{2\left| A\right| _{c}}}=\frac{D^{1/4}l^{1/2}}{%
2^{1/2}\pi ^{1/2}\mu ^{1/4}}=\frac{1}{2k_{m}}  \label{eq2.21b}
\end{equation}
We see that indeed $r_{c}$ is comparable with period of the ``strain wave''
at the temperature of the instability.

We have found in Ref. \cite{BLxy1} that the usual values of the material
coefficients involved the period $a$ of the domain structure do not depend
on temperature, and the domain width is given by the formula

\begin{equation}
a=\left( \frac{4\pi ^{3}}{21\zeta \left( 3\right) }\right) ^{1/2}\left( 
\frac{D}{\mu }\right) ^{1/4}l^{1/2}.  \label{eq2.22}
\end{equation}
This is only about $50\%$ more than the period of the ``strain wave''
appearing at the transition. In other words, the evolution of the ``wave''
into the domain structure mainly consists of changing the shape of the wave
(from sinusoidal to the piecewise domain-like) without a considerable change
of its period. Since the period does not change much, the domain width
becomes smaller than the domain wall width within the temperature interval
about $T_{c}-T_{c1}$. In a special case, when $\mu $ is anomalously small
and the film is thin enough, the period of the domain structure is much
larger than that given by Eq.~(\ref{eq2.22}) (see Ref.\cite{BLxy1}). At
these conditions one should observe a more drastic evolution of the ``strain
wave''.

\section{Strain $u_{xx}-u_{zz}$ as the order parameter}

We shall now consider the film where the soft modulus corresponds to the
strain $u_{xx}-u_{zz}$, Fig.~2. Once more, we assume that the elastic properties of
the system are those of an isotropic medium with respect to all other
strains. We shall see later that the character of the phase transition
changes drastically if the misfit strain of the form assumed in Sec.1 is
nonzero. But we first consider the case where such a misfit is absent, to
better appreciate its role.
\begin{figure}[t]
\epsfxsize=3.2in \epsffile{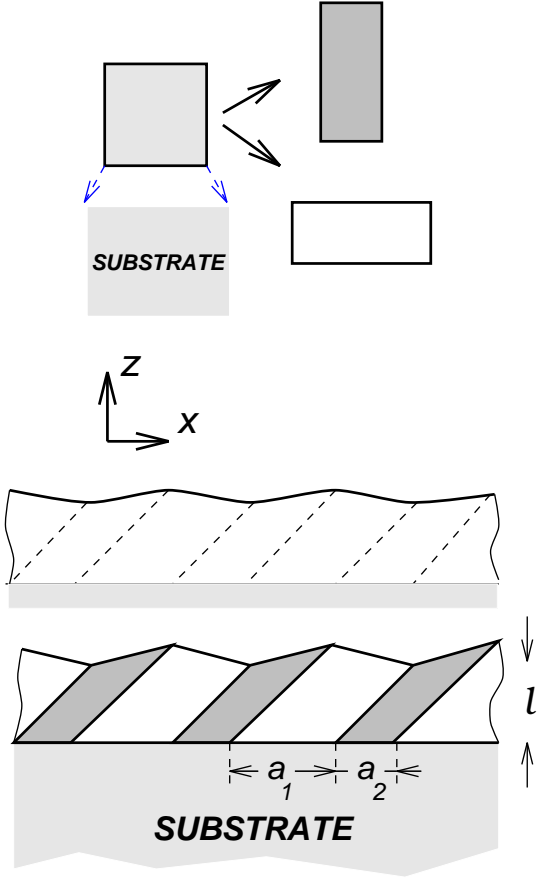}
\caption{Schematic of the $c/a/c/a$ ``strain wave'' 
and the domain structure (middle and bottom panels) in thin
epitaxial ferroelastic film with the spontaneous
strain $u^0_{xx}-u^0_{zz}$. 
The symmetric phase looses its stability at some temperature $T_{c1}<
T_c$, where $T_c$ is the critical temperature in the free film, and forms the
``strain wave'' (middle panel). 
In the presence of the ``extrinsic'' misfit
between the parent phase and the substrate (top panel)
the ``strain wave'' evolves into the $c/a/c/a$ domain structure with
the period $2a=a_1+a_2$
at low temperatures in a film of thickness $l$.
}
\label{fig:figca}
\end{figure}

\subsection{Domain pattern at the absence of extrinsic (permanent) misfit}

The same reasoning as in Sec. 2 leads to the conclusion that the soft wave
vectors make 45$^{\circ }$ angle with the film plane. With the same
reservations as in Sec.2, the Landau free energy for the film substance is
now:

\begin{eqnarray}
F &=&\int dV\Bigl[\frac{A}{2}u_{c}^{2}+\frac{D}{2}\left( \nabla u_{c}\right)
^{2} \\
&&+\frac{\lambda +\mu }{2}\left( u_{xx}+u_{zz}\right) ^{2}+2\mu u_{xz}^{2}%
\Bigr]
\end{eqnarray}
where the {\em critical} strain $u_{c}=u_{xx}-u_{zz}$, $\lambda $ and $\mu $
are the Lam\'{e} constants, and we shall consider the structures, which are
homogeneous along the $y$ axis. For the free energy density in the substrate
we put $A=\mu $ and omit the gradient term.

Therefore, for the film we have 
\begin{eqnarray}
\sigma _{xx} &=&(\lambda +\mu +A-D\nabla ^{2})u_{xx}+(\lambda +\mu
-A+D\nabla ^{2})u_{zz},  \label{eq4.21} \\
\sigma _{zz} &=&(\lambda +\mu +A-D\nabla ^{2})u_{zz}+(\lambda +\mu
-A+D\nabla ^{2})u_{xx}, \\
\sigma _{xz} &=&2\mu u_{xz},
\end{eqnarray}
while for the substrate we obtain 
\begin{eqnarray}
\sigma _{xx} &=&(\lambda +2\mu )u_{xx}+\lambda u_{zz},  \label{eq4.24} \\
\sigma _{zz} &=&(\lambda +2\mu )u_{zz}+\lambda u_{xx}.
\end{eqnarray}
We can show that there is no loss of stability with respect to homogeneous
strain. For inhomogeneous strains we have to satisfy the equations of
equilibrium 
\begin{eqnarray}
\frac{\partial \sigma _{xx}}{\partial x}+\frac{\partial \sigma _{xz}}{%
\partial z} &=&0,  \label{eq4.26} \\
\frac{\partial \sigma _{zz}}{\partial z}+\frac{\partial \sigma _{xz}}{%
\partial x} &=&0.
\end{eqnarray}
We have already mentioned in the previous section that the instability of
the symmetrical phase is signaled by appearance of the non-trivial solutions
of these equations (with the appropriate boundary conditions).

We shall reduce the equations of equilibrium to a system of ordinary
differential equations with the use of the Fourier transformation of the
functions $u_{x}\left( x,z\right) $ and $u_{z}\left( x,z\right) $%
\begin{equation}
u_{x\left( z\right) }\left( x,z\right) =\int u_{x\left( z\right) }\left(
k,z\right) \exp \left( ikx\right) dk  \label{eq4.28}
\end{equation}
We obtain the following equations for the Fourier components of the strain
vector in the film $(0<z<l)$

\begin{eqnarray}
-k^{2}\left( \alpha +1-s^{2}\right) u_{x}+ik\left( \alpha +1+s^{2}\right) 
\frac{du_{z}}{dz}+\frac{d^{2}u_{x}}{dz^{2}}+ik\frac{du_{z}}{dz} &=&0, \\
\left( \alpha +1-s^{2}\right) \frac{d^{2}u_{z}}{dz^{2}}+ik\left( \alpha
+1+s^{2}\right) \frac{du_{x}}{dz}-k^{2}u_{z}+ik\frac{du_{x}}{dz} &=&0
\end{eqnarray}
where $\alpha =\lambda /\mu ,$ $s^{2}=-\left( A+Dk^{2}\right) /\mu .$ The
general solution of this system reads 
\begin{equation}
u_{x}^{f}=\sum_{j=1}^{4}C_{j}\exp \left( \chi _{j}z\right) ,\text{\qquad }%
u_{z}^{f}=\sum_{j=1}^{4}\left( -1\right) ^{j}C_{j}\exp \left( \chi
_{j}z\right)   \label{eq4.31}
\end{equation}
where (for $0<$ $s^{2}\ll 1$) 
\begin{eqnarray}
\chi _{1,2} &=&\pm ik\left( 1-s\sqrt{\frac{2+\alpha }{1+\alpha }}\right) ,
\label{eq4.32} \\
\chi _{3,4} &=&\pm ik\left( 1+s\sqrt{\frac{2+\alpha }{1+\alpha }}\right) ,
\end{eqnarray}
and we have kept only the first two terms of the expansion of $\chi _{j}$ in
terms of $s\ll 1.$ For the substrate $(-\infty <z<0)$ the equations of
equilibrium take the form

\begin{eqnarray}
-k^{2}\left( \alpha +2\right) u_{x}+ik\alpha \frac{du_{z}}{dz}+\frac{%
d^{2}u_{x}}{dz^{2}}+ik\frac{du_{z}}{dz} &=&0,  \label{eq4.33} \\
\left( \alpha +2\right) \frac{d^{2}u_{z}}{dz^{2}}+ik\alpha \frac{du_{x}}{dz}%
-k^{2}u_{z}+ik\frac{du_{x}}{dz} &=&0.
\end{eqnarray}
The solution that vanishes at $z\rightarrow -\infty $ is

\begin{eqnarray}
u_{x}^{s} &=&\left( D_{1}+D_{2}z\right) \exp \left( kz\right) ,
\label{eq4.35} \\
u_{z}^{s} &=&-i\left[ D_{1}+D_{2}\left( z-\frac{\alpha +3}{\left( \alpha
+1\right) k}\right) \right] \exp \left( kz\right)
\end{eqnarray}
Two conditions at the free surface $\left( \sigma _{zz}=\sigma _{xz}=0\text{
at }z=l\right) $ and four conditions at the interface (continuity of $u_{x},$
$u_{z},$ $\sigma _{zz}$, $\sigma _{xz}$) provide six algebraic equations for 
$C_{j},$ $D_{j}$. This system of homogeneous equations has a nontrivial
solution if the determinant of its coefficients is zero. After a
straightforward algebra, and again assuming that $s\ll 1,$ one finds the
condition for nontrivial solution,

\begin{equation}
\cos \left( 2kls\sqrt{\frac{\alpha +2}{\alpha +1}}\right) =-\frac{2}{\left(
\alpha +1\right) ^{3}\left( \alpha +2\right) }  
\label{eq4.36}
\end{equation}
For $s^{2}<0$ one has to replace the $\cos $ by $\cosh $. It is clearly seen
that there would be no solution for $s^{2}<0$, i.e. it is the case $s^{2}>0$
that corresponds to the stability loss.

Equation $\left( \text{\ref{eq4.36}}\right) $ refers to the stability loss
for a given $k$. To find the actual stability loss one has, as in the
previous section, to find the $k$ vector that corresponds to the first
instance where the instability sets in, i.e. at the highest temperature
below $T_{c}$. To find it, one has to minimize the right hand side of the
equation 
\begin{equation}
\left| A\right| =Dk^{2}+\frac{\mu \left( \alpha +1\right) }{%
4k^{2}l^{2}\left( \alpha +2\right) }\arccos ^{2}\left( -\frac{2}{\left(
\alpha +1\right) ^{3}\left( \alpha +2\right) }\right) ,
\end{equation}
with the result 
\begin{eqnarray}
k_{m} &=&\frac{\left( \frac{\alpha +1}{\alpha +2}\right) ^{1/4}}{%
2^{1/2}l^{1/2}}\left( \frac{\mu }{D}\right) ^{1/4}\arccos ^{1/2}\left( -%
\frac{2}{\left( \alpha +1\right) ^{3}\left( \alpha +2\right) }\right)  \\
&\sim &1/\sqrt{d_{at}l}.
\label{eq:kmca}
\end{eqnarray}
We see that the condition $s^{2}\left( k_{m}\right) \sim d_{at}/l\ll 1$ is
satisfied as long as $l\gg d_{at}.$ One sees that up to a numerical factor
of order unity the result coincides with that given by Eq. (\ref{eq2.19}).
There is no essential changes if one puts $\lambda =0$ ($\alpha =0$). We
find the form of the ``deformation wave'' arising at the point of stability
loss just for this case. Assuming once more that $s^{2}\left( k_{m}\right)
\ll 1$ one finds:

\begin{eqnarray}
u_{xx}\left( x,z\right) &=&-u_{zz}\left( x,z\right)  \label{eq4.39} \\
&=&C\cos \left[ k_{m}\left( x\pm z)+\varphi \right) \right] \sin \frac{\pi z%
}{2l}, \\
u_{xz} &=&C\frac{\pi }{2k_{m}l}\sin \left[ k_{m}\left( x\pm z\right)
+\varphi \right] \cos \frac{\pi z}{2l}
\end{eqnarray}
where, once more, we obtain two possible ``strain waves'' corresponding to
two possible periodic domain systems, from which only one can actually
materialize.

We found in Ref. \cite{BLca} that if the spontaneous shear strain appearing
at the transition constitutes the only misfit strain existing in the system,
then far from the transition ($2\left| A\right| =\mu $) the period of the
domain structure is given by the formula

\begin{equation}
a_{e}\simeq l^{1/2}\left( \frac{\pi ^{3}}{7}\frac{\lambda +2\mu }{\left(
\lambda +\mu \right) \zeta \left( 3\right) }\right) ^{1/2}\left( \frac{D}{%
\mu }\right) ^{1/4}\sim d_{at}^{1/2}l^{1/2}.  
\label{eq4.40}
\end{equation}
While obtaining this formula, the surface energy of the domain walls was
estimated as $D^{1/2}u_{0}^{2}\mu ^{1/2}$, where $u_{0}$ is the spontaneous
strain in a free sample. Such an estimate is reasonable for a proper
ferroelastic. Comparing Eqs. $\left( \ref{eq4.40}\right) $ and $\left( \ref
{eq:kmca}\right) $ we see that for the case in question the period of the
domain structure far from the transition is close to its period at the
transition where the domain structure is a sinusoidal ``strain wave''. It is
quite similar to what we have seen in Sec.2.

\subsection{Effect of extrinsic misfit}

The properties of the domain pattern change substantially, if there is a
misfit strain in the temperature range corresponding to the symmetric phase.
This is usually the case, since the lattice constants of the film and the
substrate are usually different. Therefore, the film is strained (with
respect to the free film) at all temperatures. Since any misfit makes $x$-
and $z$-axes inequivalent, the strain in the film includes the ``symmetry
breaking strain'', $u_{xx}-u_{zz}$. In other words, the ``symmetric phase''
is no longer symmetric. Yet, a phase transition is possible in this system.
But, this is a transition between the single-domain and multi-domain states.
It was studied by several authors \cite
{Roytburd93,Pompe93,Pertsev95,Roytburd98,BLca}. It went unnoticed before
that this transition is qualitatively different from the discussed above: it
is of the same type as the commensurate-incommensurate transitions and/or
the transition between the homogeneous and vortex states in type II\
superconductors. In other words, there is no loss of stability of the
homogeneous (single-domain) phase at such a transition.

The theory of this transition \cite{BLca} was developed for an extreme case
which is opposite to the one considered in this paper. Indeed, the theory
assumes that the width of the domain walls is negligibly small compared to
all other length scales in the problem. We have found\cite{BLca} that at the
transition the period of the domain structure diverges, while the width of
the minority domains remains finite and equal to $\left( 2\pi l\right)
^{1/2}\left( D/\mu \right) ^{1/4}$ (in the notations of the present paper
and neglecting a logarithmic factor). This length should be larger than the
width of the domain walls $\left( D/\left| A\right| \right) ^{1/2}$, i.e.
the inequality 
\begin{equation}
\left| A\right| \gg \frac{D^{1/2}\mu ^{1/2}}{2\pi l}  \label{eq4.41}
\end{equation}
should be obeyed. Comparing with Eq.(\ref{eq2.19a}), we conclude that the
crossover between the two types of transitions occurs in the region where
the domain structure should be treated within a non-linear and non-local
theory, but this is clearly beyond the scope of the present paper.

\section{Conclusions}

We have considered two examples of proper ferroelastic transitions in thin
epitaxial films. In the first one the ferroelastic strain $u_{xy}$ breaks
the symmetry in the film plane, which was not broken by the ``extrinsic''
misfit (irrelevant to the that phase transition). There a ``usual''
(Landau-like) second order phase transition takes place with a loss of
stability of the symmetric phase and formation of sinusoidal ``strain
wave''. The period of the wave is close to the period of the domain
structure, which will evolve at lower temperatures, except for a special
case of very thin films and anomalously small non-critical elastic modulus $%
\mu $. In the second case of the ferroelastic $u_{xx}-u_{zz}$ strain the
results are similar when the ``extrinsic'' misfit (mismatch between the
lattice parameters of the symmetric phase of the film and the substrate) is
absent. If the ``extrinsic'' misfit is present, the phase transition becomes
the (topological) transition between a single- and a multi-domain states.
The crossover between the transitions of the two types remains an open issue.


\begin{references}
\bibitem{Part}  Some of the present results are briefly discussed in Ref. 
\cite{BLxy1}.
\end{references}
\end{document}